\documentclass[12pt, centerh1]{article}

\usepackage{graphicx,colonequals,bm}
\usepackage{url}
\usepackage{amsmath}
\usepackage{amsfonts}
\usepackage{amssymb}
\usepackage{url}
\usepackage{color}
\usepackage{subfigure}
\newcommand{\bx}{\boldsymbol{x}}
\newcommand{\by}{\boldsymbol{y}}
\newcommand{\bz}{\boldsymbol{z}}

\newcommand{\bA}{\boldsymbol{A}}
\newcommand{\bD}{\boldsymbol{D}}

\newcommand{\bX}{\boldsymbol{X}}
\newcommand{\bY}{\boldsymbol{Y}}

\newcommand{\bB}{\boldsymbol{B}}

\newcommand{\cL}{\mathcal{L}}

\newcommand{\EE}{\mathbb{E}}

\newcommand{\bmu}{\mbox{\boldmath $\mu$}}

\newcommand{\bchi}{\mbox{\boldmath $\chi$}}

\newcommand{\btheta}{\mbox{\boldmath $\theta$}}

\newcommand{\bpi}{\mbox{\boldmath $\pi$}}

\newcommand{\bSigma}{\mbox{\boldmath $\Sigma$}}

\begin{document}

\title{Families of Parsimonious Finite Mixtures of Regression Models}
\author{Utkarsh J.\ Dang and Paul D.\ McNicholas\thanks{Department of Mathematics \& Statistics, University of Guelph, Guelph, Ontario, N1G 2W1, Canada. E-mail: udang@uoguelph.ca.}}
\date{Department of Mathematics \& Statistics, University of Guelph}
\maketitle

\begin{abstract}
Finite mixtures of regression models offer a flexible framework for investigating heterogeneity in data with functional dependencies. These models can be conveniently used for unsupervised learning on data with clear regression relationships. We extend such models by imposing an eigen-decomposition on the multivariate error covariance matrix. By constraining parts of this decomposition, we obtain families of parsimonious mixtures of regressions and mixtures of regressions with concomitant variables. These families of models account for correlations between multiple responses. An expectation-maximization algorithm is presented for parameter estimation and performance is illustrated on simulated and real data.
\end{abstract}

\section{Introduction}
\label{sec:introduction}

Model-based clustering has become increasingly popular during the last decade. Parametric mixture models are used in model-based clustering; however, such models generally do not exploit covariates. Incorporating a regression structure can yield important insight when there is a regression relationship between some variables. Methodologies that deal with such data include finite mixtures of regressions (FMR; \cite{desarbo1988, leisch2004}) and finite mixtures of regressions with concomitant variables (FMRC; \cite{wedel2002}), supported by the popular \texttt{flexmix} package \cite{leisch2004}.

Multivariate correlated responses can be naturally integrated into such models. However, \texttt{flexmix} currently does not account for correlated response variables for both FMR and FMRC. FMR models that deal with correlated response variables have recently been proposed \cite{soffritti2011, galimberti2013}. Experimental results using these models illustrated that ignoring this correlation can lead to estimated regression coefficients with larger mean square errors and may result in a worse fit to data \cite{soffritti2011}. However, these models do not decompose the covariance structure to gain parsimony, nor do they extend the finite mixtures of regression with concomitant variables model. 

Here, FMR and FMRC are extended to deal with multiple correlated responses. Parsimonious versions of these models are developed by constraining the component covariance matrices using an eigen-decomposition in Sec. \ref{subsec:eigen}. An expectation-maximization algorithm is described in Sec. \ref{subsec:paraest}. Performance is illustrated on simulated and real data and compared to popular existing methodologies like FMR, FMRC in Sec. \ref{sec:results} with some concluding remarks in Sec. \ref{sec:discussion}. 

\section{Methodology}
\label{sec:methodology}

Let $\bX_i$ and $\bY_i$ be random vectors defined on $\Omega$ for $i = 1,\ldots,N$. Here, the response vector $\bY_i$ has values in $\mathbb{R}^d$ and the explanatory vector $\bX_i$ has values in $\mathbb{R}^p$. Then, in an FMR framework, the probability of the response $p(\by_i)$ can be decomposed as
\begin{equation}
p(\by_i| \btheta)=\sum^{G}_{g=1} p(\by_i|\bx_i,\Omega_g) \pi_{ig}, \label{FMR}
\end{equation}
where $p(\by_i|\bx_i,\Omega_g)$ is the conditional density of $\by_i$ given $\bx_i$ and $\Omega_g$ and $\pi_{ig}$ are the mixing weights, where $\pi_{ig}>0$ ($g = 1,\ldots,G$) and $\sum^G_{g=1}\pi_{ig}=1$. $\btheta$ denotes the set of all parameters. $\bY|\bX$ is assumed to be normally distributed with mean $\bmu_{\by;g}$ and covariance matrix $\bSigma_{g}$, for $g = 1,\ldots,G$. For the FMR model, $\pi_{ig}=\pi_{g}$ for $g=1,\ldots,G$ and $i=1,\ldots,N$. In addition to \eqref{FMR}, the FMRC model assumes a concomitant variable multinomial logit model for the component mixing weights, i.e.,
\begin{equation}
\bpi_{ig}(\bx,\alpha)=\frac{\exp(\alpha'_g \bx)}{\sum^{G}_{h=1} \exp(\alpha'_g \bx)}, \label{FMRCweights}
\end{equation}
with the first component as baseline. In other words, FMR only models the distribution of the $\bY|\bX$, while FMRC models both the distribution of $\bY|\bX$ and a logistic model of the concomitant variables (which may include the covariates), respectively. Note that this implies that for an FMRC model, the classification (dependent on the posterior probability) of an observation into a particular component is dependent on the covariates both through the mixing weights and $\bY|\bX$.

\subsection{Eigen decomposition of $\bSigma_{\by|\bx}$}
\label{subsec:eigen}
There are $d(d+1)/2$ free parameters in each component covariance matrix for a $d$-variate Gaussian mixture, cf.\ \eqref{FMR}. That this number increases quadratically with $d$ is undesirable for all but very low dimensional data sets. To overcome this problem, $\bSigma_{g}$ can be eigen-decomposed \cite{banfield1993} and constraints can be imposed to give a family of mixture models \cite{celeux1995}, i.e., the $g$th component covariance matrix can be written as
\begin{equation}
\bSigma_g=\lambda_g \bD_g \bA_g \bD'_g, 
\end{equation}
where $\lambda_g$ is a constant, $\bD_g$ is the orthogonal matrix of eigenvectors of $\bSigma_{g}$, and $\bA_g$ is a diagonal matrix with entries proportional to the eigenvalues of $\bSigma_g$ with the constraint $|\bA_g|=1$. Geometrically, $\lambda_g$ controls the volume, $\bD_g$, the orientation, and $\bA_g$ the shape of the $g$th component (Table \ref{gpcmtable}).

Constraining the component covariance in \eqref{FMR} leads to two families (eFMR and eFMRC, respectively) of 14 models capable of modelling the correlation between responses. This is the first time that FMR and FMRC models have been used with eigen-decomposed covariance structures, i.e., the first parsimonious families of such models.
\begin{table}[htbp]
\caption{Geometric interpretation of the eigen-decomposition of a covariance matrix.\label{gpcmtable}}
\begin{tabular*}{\textwidth}{@{\extracolsep{\fill}}p{0.3in}llllp{1.6in}}
\hline 
Name & Covariance & Volume & Shape & Orientation &  Parameters\\
\hline
EII & $\lambda I$&Equal & Spherical & -&$1$ \\
VII & $\lambda_g I$&Variable &Spherical &-&$G$\\
EEI & $\lambda \bA$&Equal &Equal &Axis-aligned&$d$ \\
VEI & $\lambda_g \bA$&Variable&Equal &Axis-aligned&$d+G-1$\\
EVI & $\lambda \bA_g$&Equal&Variable&Axis-aligned&$dG-G+1$\\
VVI & $\lambda_g \bA_g$& Variable&Variable&Axis-aligned&$dG$ \\
EEE&  $\lambda \bD\bA\bD'$& Equal &Equal &Equal &$d(d+1)/2$\\
VEE & $\lambda_g \bD \bA \bD'$ & Variable &Equal & Equal &$d(d+1)/2+G-1$ \\
EVE  & $\lambda  \bD \bA_g \bD'$&Equal & Variable &Equal &$(G-1)(p-1)+d(d+1)/2$\\
VVE & $\lambda_g \bD \bA_g \bD'$&Variable & Variable & Equal &$(G-1)p+d(d+1)/2$\\
EEV & $\lambda \bD_g\bA \bD_g'$ &Equal &Equal &Variable&$Gd(d+1)/2-(G-1)d$ \\
VEV & $\lambda_g \bD_g\bA \bD_g'$& Variable&Equal &Variable &$Gd(d+1)/2-(G-1)(d-1)$\\
EVV & $\lambda \bD_g\bA_g \bD_g'$ &Equal & Variable & Variable &$Gd(d+1)/2-(G-1)$\\
VVV & $\lambda_g \bD_g\bA_g \bD_g'$& Variable&Variable&Variable&$Gd(d+1)/2$\\
\hline
\end{tabular*}

\end{table}

\subsection{Parameter Estimation}
\label{subsec:paraest}
Parameter estimation is described here for the most unconstrained  (VVV) model from the eFMR and eFMRC families. Let ${(\bx_1,\by_1), \ldots, (\bx_N,\by_N)}$ be a sample of $N$ independent observations. The observed likelihood function under Gaussian distributional assumptions is
\begin{equation}
\label{incompletell}
L_0(\btheta| \bX, \bY)  = \prod_{i=1}^N p(\bx_i,\by_i|\btheta)  = \prod_{i=1}^N[\sum_{g=1}^G \phi_d(\by_i|\bx_i, \bchi_g) \bpi_g].
\end{equation}
Here, $\phi_d$ denotes the probability density function for a $d$ dimensional multivariate Gaussian distribution, $\bchi_g=(\bB_g, \bSigma_{g})$ refers to the parameters of the conditional distribution $p(\bY|\bX)$. Here, the covariates are supplemented by a vector of ones such that $\bB_g$ is a $(p+1)\times d$ matrix of regression intercepts and coefficients. Hence, the $(p+1,d)th$ element of $\bB_g$ denotes the regression coefficient of the $p$th predictor on the $d$th response. 

In \eqref{incompletell}, ($\bx_1, \ldots, \bx_N, \by_1, \ldots, \by_N$) are considered incomplete in the context of the EM algorithm. The complete-data are ($\bx_1, \ldots, \bx_N, \by_1, \ldots, \by_N, \bz_1, \ldots, \bz_N$), where $z_{ig}$ is a component label such that $z_{ig}=1$ if $(\bx'_i,\by'_i)'$ comes from the $g$th population and $z_{ig}=0$ otherwise. 
Therefore, the complete-data likelihood is 
\begin{equation}
\mathcal{L}_c(\btheta| \bX, \bY) = \prod_{i=1}^N \prod_{g=1}^G [\phi_{d}(\by_{i}|\bx_i, \bchi_g)  \bpi_g]^{z_{ig}},
\end{equation}
which can be decomposed as
$$\cL_c (\btheta| \bX, \bY)  = \sum_{i=1}^N \sum_{g=1}^G z_{ig} [ \log \phi_{d}(\by_{i}|\bx_i, \bchi_g) + \log \pi_g ].$$
The E-step involves calculating the expected complete data log-likelihood
\begin{equation*}
Q(\btheta,\btheta^{(k)})  = \EE_{\theta^{(k)}} \{ \cL_c(\btheta|\bX, \bY)\} =
\sum_{i=1}^N \sum_{g=1}^G  \tau_{ig}^{(k)} [ Q_{1}(\bchi_g|\btheta^{(k)}) + \log \pi_g^{(k)}],
\end{equation*}
where 
$$Q_{1}(\bchi_g|\btheta^{(k)})=\frac{1}{2}[- d \log 2\pi - \log |\bSigma^{(k)}_{g}| - (\by_i-\bB_g^{'(k)} \bx_i)' \bSigma^{(k)(-1)}_{g} (\by_i-\bB_g^{'(k)} \bx_i)],$$ 
and 
\begin{equation}\label{eqn:temp}
\tau_{ig}^{(k)} \colonequals  \EE_{\theta^{(k)}} \{Z_{ig}|\bx_i, \by_i\}= \frac{\pi_g^{(k)} \phi_d (\by_i|\bx_i, \bB_g^{(k)},\bSigma_{g}^{(k)}) } {\sum_{j=1}^G \pi_j^{(k)} \phi_d(\by_i|\bx_i, \bB_j^{(k)},\bSigma_{yj}^{(k)})}.
\end{equation}
The M-step on the $(k+1)$th iteration of the EM algorithm involves the maximization of the conditional expectation of the complete-data log-likelihood with respect to $\btheta$. The updates for $\bB_{g}^{(k+1)}$ and $\bSigma_{g}^{(k+1)}$ are
\begin{eqnarray}
\hat{\bB}_{g}^{'(k+1)} &= {\sum_{i=1}^N \tau_{ig}^{(k)}  \by_i \bx'_i}\left({\sum_{i=1}^N \tau_{ig}^{(k)}  \bx_i \bx'_i}\right)^{-1}, \label{bB_g} \\
\hat{\bSigma}_{yg}^{(k+1)} &= \frac{\sum_{i=1}^N \tau_{ig}^{(k)}  (\by_i-\hat{\bB}'_g \bx_i)(\by_i-\hat{\bB}'_g \bx_i)'}{\sum_{i=1}^N \tau_{ig}^{(k)} }. \label{bSigma_yg}
\end{eqnarray}

Note that for the VVV FMR model, the update for $\pi_{g}$ is 
\begin{equation}
\hat{\pi}_{g}^{(k+1)} = \frac{1}{N} \sum_{i=1}^N \tau_{ig}^{(k)},
\end{equation} 
and the updates for $\tau_{ig}$, $\hat{\bB}_{g}$, and $\hat{\bSigma}_{yg}$ are updated via \eqref{eqn:temp}, \eqref{bB_g}, and \eqref{bSigma_yg}, respectively. For the VVV FMRC model, the algorithm consists of updating $\hat{\bpi}_{ig}$, $\tau_{ig}$, $\hat{\bB}_{g}'$, and $\hat{\bSigma}_{yg}$ via \eqref{FMRCweights}, \eqref{eqn:temp}, \eqref{bB_g}, and \eqref{bSigma_yg}, respectively. For the FMRC model, note that $\pi_{ig}$ and $\pi_{ij}$ are used in place of $\pi_{g}$ and $\pi_{j}$, respectively in \eqref{eqn:temp}. Parameter estimates for the concomitant parameters in \eqref{FMRCweights} are estimated using function \texttt{multinom} from the \texttt{nnet} package \cite{venables2002} with the dependent variables given by the \textit{a~posteriori} probability estimates $\tau_{ig}$. For the other eFMR and eFMRC models, the M-step updates vary only with respect to the component covariance matrix $\bSigma_{g}$ and are similar to those in \cite{celeux1995}.

\subsection{Model selection and initialization}
\label{modelselection}
For choosing a `best' fitted model among a family of models, a model selection criterion like the BIC is typically used \cite{dasgupta1998}: 
$$\text{BIC}=2l(\hat{\theta})-m\log{N},$$
 where $l(\hat{\theta})$ is the maximized log-likelihood and $m$ is the number of free parameters. Even though mixture models generally do not satisfy the regularity conditions for the asymptotic approximation used in the development of the BIC \cite{keribin2000}, it has performed quite well in practice and has been used extensively \cite{fraley2002}. 

Note that the EM algorithm can be overly dependent on starting values. Singularities and convergence to local maxima are also well documented \cite{titterington1985}. Initializing the EM algorithm multiple times using k-means \cite{hartigan1979} or random initializations can alleviate some of these issues. Specifically, our EM algorithms are each initialized from five starting values, where the first four are random and the other uses $k$-means clustering.

\subsection{Convergence criterion and performance assessment}

An Aitken acceleration-based stopping criterion is used to determine convergence of our EM algorithms. This criterion is at least as strict as lack of progress in likelihood \cite{mcnicholas2010}. The Aitken acceleration at iteration $k$ is 
$$a^{(k)}=[{l^{(k+1)}-l^{(k)}}]/[{l^{(k)}-l^{(k-1)}}],$$ 
where $l^{(k)}$is the log-likelihood value at iteration $k$. An asymptotic estimate of the log-likelihood at iteration $k+1$ is given by \cite{bohning1994} as
$$l_{A}^{(k+1)}=l^{(k)}+[l^{(k+1)}-l^{(k)}]/[1-a^{(k)}],$$ 
and the EM is algorithm is stopped when $l_A^{(k+1)}-l^{k}<\epsilon$ \cite{lindsay1995}.

The adjusted Rand index (ARI; \cite{hubert1985}) is used to compare predicted and true classifications. The ARI calculates the agreement between true and estimated classification by correcting the Rand index \cite{rand1971} to account for chance. An ARI of 1 corresponds to perfect clustering, whereas the expected value of the ARI under random classification is $0$.

\section{Results}
\label{sec:results}
Performance of the proposed models is illustrated on simulated and real data. To facilitate comparison of the performance of the algorithms, the \texttt{flexmix} FMR and FMRC models are initialized with the same set of values as the eFMR and eFMRC models (Sec. \ref{modelselection}). We used the {\tt mixture} package \cite{mixture} for the M-step updates for the 14 covariance structures.

\subsection{Simulated Data}
Data were generated from a two-component model with 275 observations in total. A binomial model with $\pi_1=0.45$ was used to determine the component sizes. Here, the three-dimensional response was generated using an EEE covariance structure. Three covariates were generated. For the first component, one came from a uniform distribution with support $[0,3]$ and the others from a two-dimensional Gaussian distribution with mean $\bmu_{x1}=(0,1)$. Covariates for the second group were generated from a uniform distribution with support $[-1,5]$ and a two-dimensional Gaussian distribution with mean $\bmu_{x2}=(-3,3)$. The covariance matrices of the normally distributed covariates for the two groups were $$\left(\begin{matrix}
1&0.8\\ 
0.8&1.2
\end{matrix} \right)$$ and 
$$\left(\begin{matrix}
1.2&0.4\\ 
0.4&1
\end{matrix} \right),$$ 
respectively. The regression coefficient matrices used for the two groups were $$\left(\begin{matrix}
-1.9 & 0.4 & -1.2 & -3\\ 
0 & -0.4 & 0.8 & -2\\
-1 & 0.7 & 0.3 & 1
\end{matrix} \right)'$$ and 
$$\left(\begin{matrix}
2.5 & -0.5 & 1 & -4\\ 
2.3 & -1.3 & 1.9 &2\\
1 & -2.7 & -2.3&-1.3
\end{matrix} \right)',$$ 
respectively. Lastly, the error matrices (with mean $\boldsymbol{0}$) for the two groups using a EEE covariance structure were $$\left(\begin{matrix}
1.31 & 0.77 & 0.68 \\ 
0.77 & 1.70 & 1.06 \\ 
0.68 & 1.06 & 1.90 
\end{matrix} \right).$$ 
This corresponds to $\lambda_{1}=\lambda_{2}=1.25$, $$D_1=D_2=\left(\begin{matrix}
-0.45 & 0.72 & 0.53 \\ 
-0.62 & 0.18 & -0.76 \\ 
-0.65 & -0.67 & 0.36
\end{matrix} \right),$$ and $A_1=A_2$ (diagonal matrices) with entries $(2.7, 0.7, 1/(2.7\times 0.7))$.

A total of 50 samples were generated in {\sf R} \cite{R2013} and run for $G=1,\ldots,4$. The parameter estimates for the selected model using the eFMR and eFMRC families were quite close to the generating values (results not shown). Summary statistics for the selected models are given in Table \ref{simsum}. Clearly, the eFMR and eFMRC families perform much better. The models selected from both the eFMR and eFMRC families yielded higher average ARI and log-likelihood values. Furthermore, these models also yielded superior BIC values and estimated fewer parameters on average. Note that the range of the number of parameters fitted for the FMR and FMRC models is quite wide, implying that these models are overestimating the number of components. Specifically, the FMR and FMRC models overestimate the number of components 40 and 35 times, respectively. On the other hand, the selected eFMRC models always fitted the right number of components. The selected eFMR models fitted the right number of components 49 out of 50 times. Therefore, in contrast to the \texttt{flexmix} FMR and FMRC models, the proposed parsimonious models deal with correlations between the response variables.
\begin{table}[htbp]
\caption{Simulation study.} \label{simsum}
\begin{tabular}{lcccc}
\hline
Statistic & FMR & FMRC & eFMR & eFMRC\\
  \hline
ARI & 0.64 (0.43, 1.00) & 0.70 (0.49, 1) & 0.96 (0.86, 1.00) & 1 (0.96, 1)\\ 
$\cL_0$ & -1481 (-1538, -1389) & -1300 (-1376, -1201) & -1425 (-1476, -1381) & -1253 (-1293, -1209) \\ 
BIC & -3220 (-3332, -3130) & -2894 (-2995, 2779) & -3029 (-3127, -2937) & -2696 (-2778, -2609)\\
df & 47 (31, 63) & 53 (34, 72) & 31 (31, 46) & 34 (34, 35)\\
\hline
\end{tabular}

Values denote the medians (rounded to 2 decimals) with the ranges of the estimated statistics in parentheses. Here, $\cL_0$ refers to the maximized log-likelihood value.
\end{table}

\subsection{Crabs Data}
The crabs data set contains five morphological measurements on 200 crabs, split evenly between both sexes and two colours (blue and orange) of the species \emph{Leptograpsus variegatus}. These data were originally introduced in \cite{campbell1974} and are available as part of the {\tt{MASS}} package \cite{venables2002} in {\sf R}. The data are famous for having highly correlated measurements on width of frontal region just anterior to frontal tebercles (FL), width of posterior region (RW), carapace length (CL), carapace width (CW), and body depth (BD). The variables CW, FL, and RW reflect width measurements and were taken to be the response variables, with CL and BD as the predictor variables. Based on the two binary variables, sex and colour, there are four known groups in this data. Our algorithms were run for $G=1,\ldots,9$ (Table \ref{crabstable}).
\begin{table}[htbp]
\caption{Model performance comparison for crabs data. \label{crabstable}}
\begin{tabular*}{\textwidth}{@{\extracolsep{\fill}}llllllll}
\hline 
Algorithm & Model & $G$ & BIC & ARI & Parameters\\
\hline
FMR &  & 2 & -1178.45 & 0.40 & 25 \\
FMRC &  & 4 & -1104.96 & 0.81 & 57\\
eFMR & VVI & 2 & -1178.38 & 0.40 & 25 \\
eFMRC & VEE & 4 & -1069.36 & 0.84 & 54\\
\hline
\end{tabular*}
\end{table}

The selected eFMR model is a two-component VVI model with an ARI of 0.40. Because the VVI model assumes independence between the response variables, that is equivalent to the \texttt{flexmix} FMR model and unsurprisingly, FMR chooses a two-component model with an ARI of 0.40 (Table~\ref{classcrabs}). Note that the estimated classification from the selected two-component eFMR model leads to good separation between sexes. If the class membership agreement is estimated based on only the sexes of the crabs, an ARI of 0.81 is achieved. FMRC did well, picking a four-component model (Table \ref{classcrabs}). However, the selected eFMRC model (VEE) also has four components with a higher ARI of 0.84, while also being more parsimonious than the \texttt{flexmix} FMRC model.
\begin{table}[htbp]
\caption{True and estimated components for the crabs data. \label{classcrabs}}
\begin{tabular*}{1.0\textwidth}{@{\extracolsep{\fill}}llllllllllllllll}
            \hline
           & \multicolumn{4}{c}{FMRC}&& \multicolumn{4}{c}{eFMRC} && \multicolumn{2}{c}{FMR} && \multicolumn{2}{c}{eFMR}\\
		&1&2&3&4&&1&2&3&4&&1&2&&1&2\\
		\cline{2-5}\cline{7-10}\cline{12-13}\cline{15-16}
                 BM& 38& 12 & & && 40 &10 & & &&46&4&&46&4\\
                 BF& &48 & &2 &&  & 49 & &1 &&4 &46&&4&46\\
                 OM& & & 50 & && & &50 & && 50& &&50& \\
                 OF& & &2 &48&& & &2 &48&& 2&48 &&2&48\\

			\hline
\end{tabular*}\\
``B", ``O", ``M", ``F" refer to blue, orange, male and female, respectively.
\end{table}

\section{Discussion}
\label{sec:discussion}
Families of parsimonious multivariate response FMR and FMRC models that can handle correlated response variables were proposed and illustrated. In a model-based clustering context, we showed that both eFMR and eFMRC families perform as well as or better than the \texttt{flexmix} FMR and FMRC models.  Computationally, the algorithms were quite stable. However, to prevent fitting issues, the component sizes were computed before each M-step and a preset minimum size of the clusters was used [cf.\ \cite{leisch2004}]. For heavier tailed data, more robust distributions like the multivariate student-$t$ distribution may be employed. Because the number of regression intercepts and coefficients estimated, i.e., $Gd(p+1)$, can also increase quickly, more parsimonious models can be achieved using variable selection.

\section*{Acknowledgements}
This work is supported by a Alexander Graham Bell Canada Graduate Scholarship (CGS-D; Dang), as well as a Discovery Grant from the Natural Sciences and Engineering Research Council (NSERC) of Canada (McNicholas).


\end{document}